%% file: Paper-0554.tex
\documentclass[runningheads]{llncs}
\usepackage[T1]{fontenc}
\usepackage{graphicx,verbatim}
\usepackage{amsmath,amsfonts}
\usepackage{algorithmic}
\usepackage{algorithm}
\usepackage{array}
\usepackage[caption=false,font=normalsize,labelfont=sf,textfont=sf]{subfig}
\usepackage[colorlinks,urlcolor=blue,linkcolor=blue,citecolor=blue]{hyperref}
\usepackage{textcomp}
\usepackage{stfloats}
\usepackage{url}
\usepackage{verbatim}
\usepackage{cite}

\usepackage{amsfonts} 
\usepackage{amssymb}
\usepackage{booktabs}
\usepackage{multirow}
\usepackage{graphicx}
\usepackage{float}
\usepackage{array}
\usepackage{amsmath}
\usepackage{marvosym}
\usepackage{hyperref}
\usepackage{epstopdf}
\usepackage{soul}
\usepackage{color, xcolor}
\usepackage{algorithmic}
\usepackage{algorithm}
\usepackage{bbm}
\usepackage{threeparttable}
\usepackage{tabularx}
\usepackage{booktabs}
\usepackage{makecell}
\usepackage{wrapfig}

\newcolumntype{P}[1]{>{\centering\arraybackslash}p{#1}}
\newcolumntype{M}[1]{>{\centering\arraybackslash}m{#1}}

\makeatletter
\renewcommand\normalsize{%
\@setfontsize\normalsize\@xpt\@xiipt
\abovedisplayskip 5\p@ \@plus2\p@ \@minus5\p@
\abovedisplayshortskip \z@ \@plus3\p@
\belowdisplayshortskip 6\p@ \@plus3\p@ \@minus3\p@
\belowdisplayskip \abovedisplayskip
\let\@listi\@listI}
\makeatother

\begin{document}
\title{SAGCNet: Spatial-Aware Graph Completion Network for Missing Slice Imputation in Population CMR Imaging}
\titlerunning{SAGCNet}

\author{Junkai Liu\inst{1}
\and Nay Aung\inst{4,5}
\and Theodoros N. Arvanitis\inst{1}
\and Stefan K. Piechnik\inst{2}
\and Joao A C Lima\inst{3}
\and Steffen E. Petersen\inst{4,5}
\and Le Zhang\inst{1,4}\textsuperscript{(\Letter)}
}


\authorrunning{J. Liu et al.}
\institute{School of Engineering, College of Engineering and Physical Sciences,\\ University of Birmingham, Birmingham, UK \\
\and Oxford Center for Clinical Magnetic Resonance Research (OCMR),\\ Division of Cardiovascular Medicine, John Radcliffe Hospital,\\University of Oxford, Oxford, UK\\
\and Division of Cardiology, Johns Hopkins University School of Medicine, \\Baltimore, Maryland, USA\\
\and William Harvey Research Institute, NIHR Barts Biomedical Research Centre, Queen Mary University London, London, UK\\
\and Barts Heart Centre, St Bartholomew’s Hospital, Barts Health NHS Trust, West Smithfield, London, UK\\
     \email{jxl1920@student.bham.ac.uk; l.zhang.16@bham.ac.uk}
}


\maketitle              

\input{"Sections/0_abstract.tex"}
\input{"Sections/1_introduction.tex"}
\input{"Sections/2_methodology.tex"}
\input{"Sections/3_experiments.tex"}

\input{"Sections/4_conclusion.tex"}

\section*{Disclosure of Interests}
The authors have no competing interests in the paper.

\bibliographystyle{splncs04}
\bibliography{Paper-0554}

\end{document}

%% file: Sections/0_abstract.tex
\begin{abstract}
Magnetic resonance imaging (MRI) provides detailed soft-tissue characteristics that assist in disease diagnosis and screening. However, the accuracy of clinical practice is often hindered by missing or unusable slices due to various factors. Volumetric MRI synthesis methods have been developed to address this issue by imputing missing slices from available ones. The inherent 3D nature of volumetric MRI data, such as cardiac magnetic resonance (CMR), poses significant challenges for missing slice imputation approaches, including (1) the difficulty of modeling local inter-slice correlations and dependencies of volumetric slices, and (2) the limited exploration of crucial 3D spatial information and global context. In this study, to mitigate these issues, we present \textbf{S}patial-\textbf{A}ware \textbf{G}raph \textbf{C}ompletion \textbf{N}etwork \textbf{(SAGCNet)} to overcome the dependency on complete volumetric data, featuring two main innovations: (1) a volumetric slice graph completion module that incorporates the inter-slice relationships into a graph structure, and (2) a volumetric spatial adapter component that enables our model to effectively capture and utilize various forms of 3D spatial context. Extensive experiments on cardiac MRI datasets demonstrate that SAGCNet is capable of synthesizing absent CMR slices, outperforming competitive state-of-the-art MRI synthesis methods both quantitatively and qualitatively. Notably, our model maintains superior performance even with limited slice data. Code is available at \url{https://github.com/JK-Liu7/SAGCNet}.


\keywords{Medical image synthesis \and Incomplete graph \and Cardiac MRI}

\end{abstract}

%% file: Sections/1_introduction.tex
\section{Introduction}
Magnetic resonance imaging (MRI) provides significant insights into tissue and anatomical characteristics and is widely used in medical research and clinical diagnosis. Nonetheless, missing slice is a common issue for volumetric MRI data, particularly in cardiac magnetic resonance (CMR) volumes, within clinical applications and practice, caused by factors such as excessive scanning times, image deterioration, motion artifacts, and disparate acquisition techniques \cite{zhang2024automatic}. Consequently, the development of a unified and effective approach for imputing missing slices using available data is critically needed \cite{ZHANG2024102381}. 
\par Missing data imputation is a general method for tackling the incomplete volumetric data problem, employing medical slice synthesis techniques to generate missing slices from available 3D CMR images \cite{XIA2021101812} \cite{zhang2019missing} \cite{zhang2019unsupervised}. Deep learning-based medical slice synthesis, using algorithm such as Convolutional neural network (CNN) \cite{YURT2021101944} and transformer \cite{10081095}, has demonstrated notable progress and become an emerging research topic. Even though multi-modal MRI synthesis for missing modality imputation has made significant strides in recent years \cite{10589432}, these methods are not specifically devised for the missing slice imputation task, resulting in a lack of flexibility in handling arbitrary missing scenarios. 

\par Below, we highlight the following two key challenges that need be addressed for missing slice imputation task: 
\textit{(a) How to model inter-slice correlations and dependencies explicitly to capture local interactions between slices?} The inter-slice correlations focus on the interaction between adjacent slices along the through-plane in each volume, which is essential for learning discriminative and informative slice representations \cite{Fang_2022_CVPR}. Additionally, the internal dependencies among different slices contribute to describing detailed anatomical structures and lesions, which should be carefully modeled to fully explore and extract high-level and hierarchical concepts \cite{10183842}. Nevertheless, these inter-slice dependencies are often overlooked due to their complex interdependence characteristics, making them difficult to estimate and model. 
\textit{(b) How to extract the crucial 3D spatial information in medical volumetric data to fully describe the global spatial context?} CMRs are inherently volumetric and three dimensional, making it challenging for existing methods to handle their depth-wise properties \cite{Peng_2020_CVPR}. Moreover, directly applying simple 3D models to high-resolution volumes is infeasible due to the significant computational memory, and data acquisition \cite{10.1007/978-3-031-72104-5_63}. Therefore, for 3D CMR synthesis, inherent isotropic 3D spatial information must be carefully and comprehensively considered to effectively learn 3D volumetric spatial patterns and capture structural and fine-grained details. 

\begin{figure*}[t!]
\centerline{\includegraphics[width=30pc]{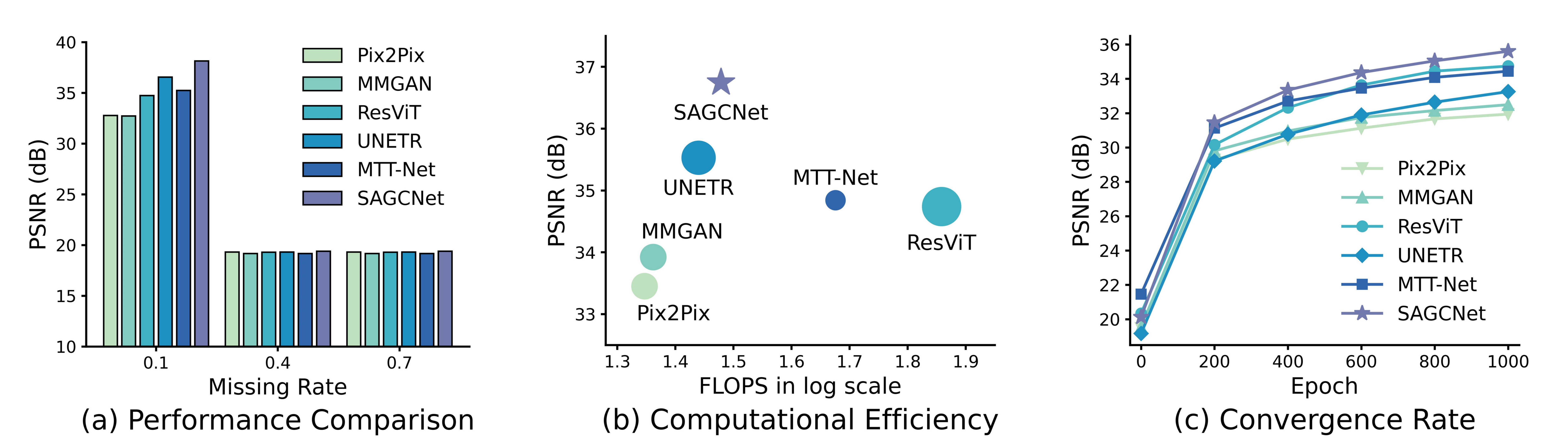}}
\caption{(a) Performance comparison of SAGCNet and baselines under missing rate of 0.1, 0.4 and 0.7. (b) PSNR vs. floating point operations (FLOPs). The area of each blob denotes the number of model parameters. (c) Convergence rate of SAGCNet and baselines under missing rate of 0.1. All experiments are conducted on the UK Biobank dataset. More details are presented in the experiments section.}
\label{Fig1}
\end{figure*}

\par To address the aforementioned challenges, we propose \textbf{S}patial-\textbf{A}ware \textbf{G}raph \textbf{C}ompletion \textbf{N}etwork \textbf{(SAGCNet)} for missing slice imputation. SAGCNet leverages graph structure to explicitly model inter-slice relationships and incorporates adapters to learn 3D spatial-related knowledge and characteristics within volumes. SAGCNet is intricately designed to balance computational cost and performance, achieving lightweight and efficient CMR synthesis with \textbf{moderate computational cost} and \textbf{rapid convergence rate}, as depicted in Fig. \ref{Fig1}. 
\par Our main contributions are as follows: 
\textbf{(1)} A unified missing slice imputation framework, SAGCNet, is proposed to effectively synthesize CMR images for arbitrary missing scenarios. 
\textbf{(2)} We propose a volumetric slice graph completion (VSGC) module, employing graphs to capture inter-slice correlations. To the best of our knowledge, our work is the first to leverage graph-based modeling at slice level for medical image synthesis. 
\textbf{(3)} We introduce a simple yet effective volumetric spatial adapter (VSA) to preserve 3D volumetric spatial information, enabling our model to extract crucial volumetric insights and thus be spatial-aware.
\textbf{(4)} Experimental results on three datasets demonstrate the quantitative and qualitative superiority of SAGCNet under various missing rates.

%% file: Sections/2_methodology.tex
\section{Methodology}
\subsection{Problem Formulation and Model Overview}
\begin{figure*}[t!]
\centerline{\includegraphics[width=30pc]{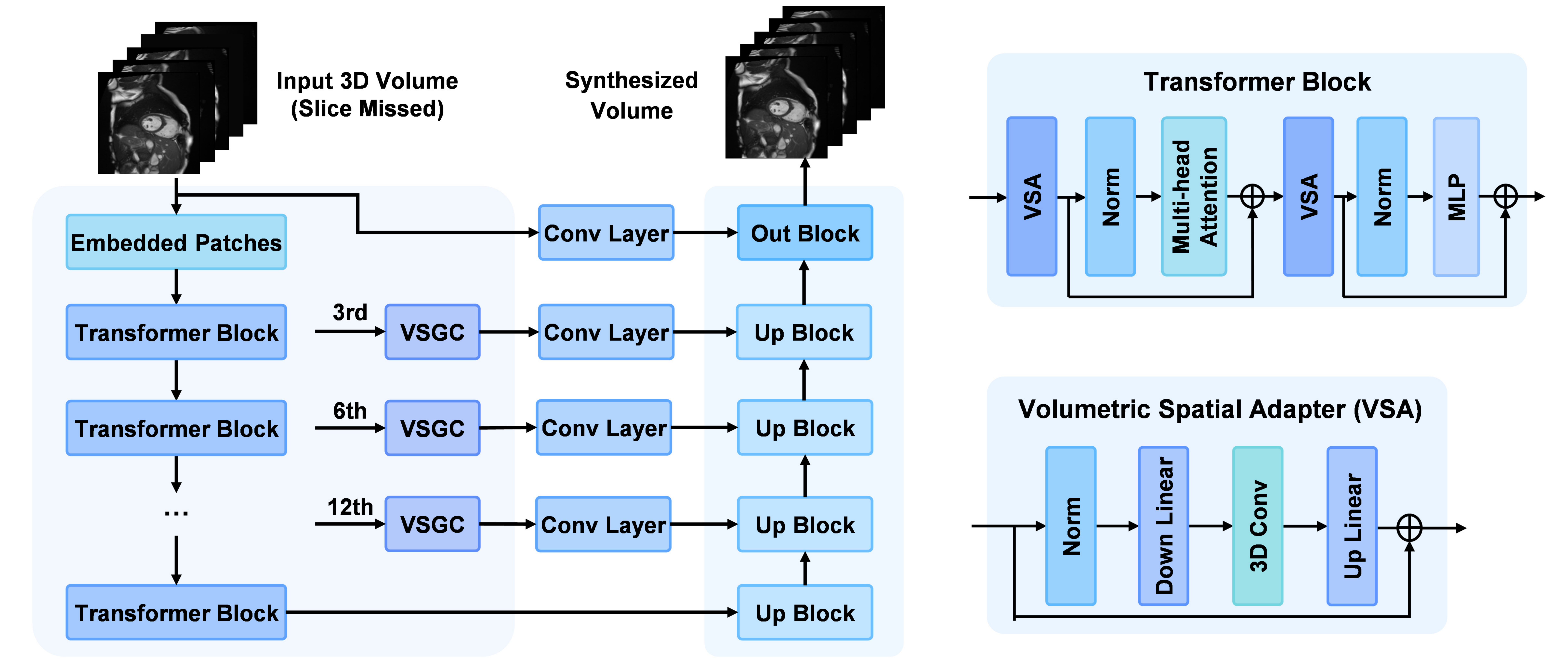}}
\caption{The overview of our proposed SAGCNet.}
\label{Fig2}
\end{figure*}

Given a volume with randomly missing slices, our aim is to construct a unified and robust framework to handle arbitrary missing slice scenarios, i.e., various missing positions and numbers, which simulates the practical clinical scenario. Mathematically, consider the given incomplete volume $\mathbf{V} \in \mathbb R^{N \times H \times W}$ with available slices $\mathbf{V}_a \in \mathbb R^{M \times H \times W}$ and missing slices $\mathbf{V}_c \in \mathbb R^{P \times H \times W}$, where $M$ and $P$ denote the numbers of available and missing slices, respectively, and $N=M+P$. SAGCNet reasons about the missing slice position automatedly and synthesizes the complete volume as output according to the input available slices at once. The overall architecture of SAGCNet is depicted in Fig. \ref{Fig2}. SAGCNet utilizes modified UNETR \cite{Hatamizadeh_2022_WACV} with VSA as visual encoder to extract hierarchical features. Moreover, VSGC blocks are incorporated into the backbone encoder to fully exploit inter-slice relationships from multi-views. Ultimately, following the U-shaped network, the CNN-based decoders with skip connections are employed to reconstruct and synthesize images. Next, we will elaborate the detailed information of VSA and VSGC modules. 

\subsection{Volumetric Spatial Adapter Module}
To bridge the gap between 2D images and volumetric medical data, we devise a series of VSA modules integrated into each transformer block, empowering SAGCNet to learn spatial information inherent in 3D volumetric medical data \cite{NEURIPS2022_a92e9165}. Specifically, as illustrated in Fig. \ref{Fig2}, each VSA can be represented as 
\begin{equation}
VSA(\mathbf{H})=\mathbf{H}+\sigma\left(\operatorname{Conv} 3 \mathrm{D}\left(\operatorname{Norm}(\mathbf{H}) \cdot \mathbf{W}_{\text {down }}\right)\right) \mathbf{W}_{\text {up}}
\end{equation}
where $\mathbf{H}$ denotes the original feature representation, $\mathbf{W}_{\text {down}}$ and $\mathbf{W}_{\text {down}}$ represent the down- and up-projection layer respectively, $\operatorname{Conv} 3 \mathrm{D}$ indicates the 3D depth-wise convolutional layer. The 3D convolutional layer serves as a core module, aiming to extract valuable volumetric information. The down-projection layer reduces the original dimensionality, thereby lowering the number of parameters and making the VSA more flexible and lightweight. For each Transformer block, two VSAs are plugged before and after the multi-head self-attention module to produce better performance in practice.

\subsection{Volumetric Slice Graph Completion}

Leveraging the strong representation learning capability of Graph Neural Networks (GNNs) in complex relation modeling, graphs can effectively depict the spatial correlations of inter-slice properties through the message propagation pipeline \cite{10.1145/3580305.3599410, PENG2024102024}. To this end, we propose to model volume data as graphs and establish a multi-view graph completion network to discover node interactions within each volume and extract inter-slice dependencies. 

\begin{figure*}[t!]
\centerline{\includegraphics[width=29pc]{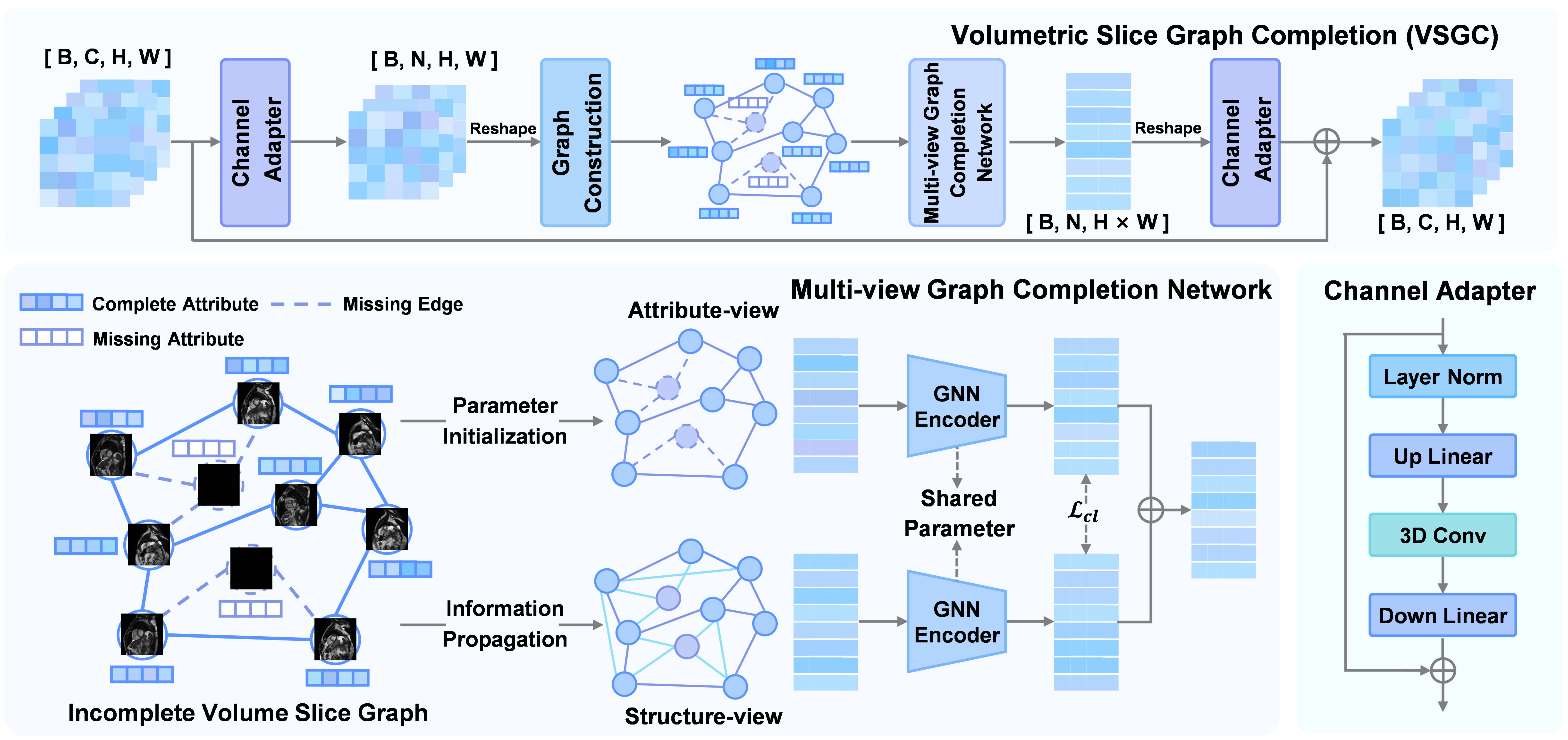}}
\caption{An illustration of the VSGC module.}
\label{Fig3}
\end{figure*}

\textbf{Graph Construction.} After obtaining feature representations from the transformer blocks, we first employ a channel adapter to construct the node features, as shown in Fig. \ref{Fig3}. Next, the k-Nearest Neighbors (kNN) algorithm is used to construct the incomplete volume slice graph $\mathcal{G}=\{\mathcal{V}, \mathcal{E}\}$, where V and E represent the node (slice) and edge (inter-slice correlation) set, respectively. Assume the slice number of a given volume, (i.e., node number) is n. Let $\mathbf{X} \in \mathbb R^{n \times d}$ and $\mathbf{A} \in \mathbb R^{n \times n}$ denote the original node attribute and adjacency matrix, respectively. Notably, nodes corresponding to missing slices are treated as nodes with missing attributes, and their associated edges are also considered incomplete.

\textbf{Multi-view Graph Completion Network.} Drawing inspiration from previous work \cite{Huo_Jin_Li_He_Yang_Wu_2023}, we devise a multi-view graph completion network to impute missing graph data at both attribute and structure levels. The attribute-view learning process, with parameterized completion, focus on extracting essential information from incomplete attributes while minimizing the impact of the incomplete graph structure on learning. Meanwhile, the structure-view learning process is designed to effectively model the incomplete graph structure without being hindered by missing node attributes. 
\par Specifically, for attribute-view imputation, we use learnable neural network parameters to initialize node attributes, reducing noise in the original attribute matrix through iterative refinement. This strategy generates the augmented graph $\mathcal{G}^a=(\mathbf{X}^a, \mathbf{A})$. Regarding structure-view completion, we utilize personalized PageRank \cite{gasteiger2018predict} to propagate information and enhance its diversity, producing $\mathcal{G}^s=(\mathbf{X}^s, \mathbf{A}^s)$, which is crucial for learning comprehensive and discriminative node representations under conditions where partial edges are missing. We choose Graph attention network \cite{velivckovic2017graph} (GAT) as the GNN encoder due to its expressive power, as follows: 
\begin{equation}
\mathbf{Z}^{a}=f_{G A T}\left(\mathcal{G}^{a}\right), \mathbf{Z}^{s}=f_{G A T}\left(\mathcal{G}^{s}\right),
\end{equation}
\par Afterwards, to bridge the semantic gap between different views resulting from graph incompleteness and to achieve consistent representations, the contrastive learning paradigm is applied to maximum the mutual information:
\begin{equation}
\mathcal{L}_{c l}^{v}=-\frac{1}{2 n} \sum_{i=1}^{n}\left(\log \frac{\varphi\left(\mathbf{z}_{i}^{a}, \mathbf{z}_{i}^{s}\right)}{\sum_{j \neq i}^{n} \varphi\left(\mathbf{z}_{i}^{a}, \mathbf{z}_{j}^{s}\right)}+\log \frac{\varphi\left(\mathbf{z}_{i}^{s}, \mathbf{z}_{i}^{a}\right)}{\sum_{j \neq i}^{n} \varphi\left(\mathbf{z}_{i}^{s}, \mathbf{z}_{j}^{a}\right)}\right),
\end{equation}
where $\varphi\left(a, b\right)=e^{sim(a,b)/\tau}$, $sim(\cdot,\cdot)$ is the cosine similarity function, $\tau$ is the temperature parameter. Finally, the fused embeddings from the two views are passed through another channel adapter with a residue connection to transform dimensionality for subsequent operation. The VSGC components are inserted after the 3rd, 6th, and 12th transformer blocks, respectively. 

\textbf{Loss Function.} The overall loss function of SAGCNet is defined as 
\begin{equation}
\mathcal L = \lambda_{r} \mathcal L_{rec} + \lambda_{s} \mathcal L_{syn} + \lambda_{cl} \mathcal L_{cl}   
\end{equation}
where $\mathcal L_{rec}$, $\mathcal L_{syn}$, and $\mathcal L_{cl}$ denote the reconstruction loss, synthesis loss and contrastive loss, respectively, with $\lambda_{r}$, $\lambda_{s}$, and $\lambda_{cl}$ are trade-off parameters for balance. $\mathcal L_{rec}$, $\mathcal L_{syn}$ consist of L1 loss and perceptual loss. 

%% file: Sections/3_experiments.tex
\section{Experiments}
\noindent \textbf{Datasets.} We validate the performance of the proposed SAGCNet on three cardiac MRI datasets: UK Biobank (UKBB) \cite{PETERSEN20168}, Multi-Ethnic Study of Atherosclerosis (MESA) \cite{10.5665/sleep.4732} and Automatic Cardiac Diagnosis Challenge (ACDC) \cite{zhang2025diffuseg}. The UKBB dataset we utilized contains CMR scans from 600 patients, each with 50 temporal phases. The MESA dataset consists of 299 CMR volumes, with the number of slices per volume ranging from 6 to 14. The ACDC dataset is a publicly available resource containing CMR images from 100 patients, with 6 to 21 slices per volume. We divide all datasets into training and test sets in the ratio of 8:2. During preprocessing, each 2D slice image is resized to 256 × 256. Zero-padding is applied for all volumes in the through-plane direction to ensure the fixed input size. Min-max normalization is employed to scale the intensity range of all images to [-1, 1]. To enhance data diversity, random flipping and rotation are applied as data augmentation. 

\textbf{Implementation Details.} The missing rate $\eta$ is defined as $\eta=P/N$, which is fixed during training and inference stages. For training phase, the missing slices are randomly sampled based on $\eta$ at each iteration to ensure that our model is robust to arbitrary missing scenario. For model hyperparameters, we set the number of kNN neighbors to 3 and the number of GAT layers to 2. The hyperparameter $\tau$ is set empirically to 0.8. Additionally, $\lambda_r$, $\lambda_s$, and $\lambda_{cl}$ are set to 5, 20, and 0.001, respectively. All experiments are implemented on an Nvidia A100 GPU. We train SAGCNet for 2000 epochs using the Adam optimizer with an initial learning rate of 1e-4. The cosine annealing scheduler is applied to decay the learning rate to 5e-6, and the training batch size is set to 8. 

\textbf{Baselines and Evaluation Metrics.} We compare the proposed SAGCNet with several state-of-the-art (SOTA) image synthesis methods as baselines, including Pix2Pix \cite{Isola_2017_CVPR}, MMGAN \cite{8859286}, ResViT \cite{9758823}, UNETR \cite{Hatamizadeh_2022_WACV}, and MTT-Net \cite{10268458}. To quantitatively evaluate the performance of SAGCNet and baselines, three commonly-used evaluation metrics are adopted, including mean absolute error (MAE), peak signal-to-noise ratio (PSNR), and structural similarity (SSIM).

\textbf{Quantitative Results.} Table \ref{Table1} presents the quantitative results of our proposed SAGCNet and baseline models on three datasets under the missing rate $\eta$=0.1, 0.4 and 0.7. The experimental results demonstrate that our SAGCNet consistently outperforms SOTA baselines across various missing rate settings, highlighting its superiority and effectiveness in synthesizing high-fidelity CMR images. Notably, SAGCNet shows greater performance gains under low missing rate settings (e.g., $\eta$=0.1), with improvements ranging from 1.02\%\ to 4.37\%\ in PSNR. Furthermore, even in high missing rate scenarios  (e.g., $\eta$=0.4 and 0.7), SAGCNet also surpasses competitive baselines across three datasets, underscoring its robustness and adaptability to real-world slice missing patterns.

\begin{table}[!h]
  \centering
  \setlength{\tabcolsep}{0.9mm}
  \renewcommand{\arraystretch}{0.6}
  \footnotesize
  \caption{Quantitative performance comparison of different methods on three datasets under three missing rate configurations. The best performance is in \textbf{bold}.}
    \begin{tabular}{clcc||cc||cc}
    \toprule
    \multirow{3}[2]{*}{\textbf{Dataset}} & \multirow{3}[2]{*}{\textbf{Method}} & \multicolumn{6}{c}{\textbf{Missing Rate}} \\
          & \multicolumn{1}{c}{} & \multicolumn{2}{c||}{0.1} & \multicolumn{2}{c||}{0.4} & \multicolumn{2}{c}{0.7} \\
          & \multicolumn{1}{c}{} & \multicolumn{1}{c}{PSNR $\uparrow$} & \multicolumn{1}{c||}{SSIM $\uparrow$} & \multicolumn{1}{c}{PSNR $\uparrow$} & \multicolumn{1}{c||}{SSIM $\uparrow$} & \multicolumn{1}{c}{PSNR $\uparrow$} & \multicolumn{1}{c}{SSIM $\uparrow$} \\
    \midrule
    \multicolumn{1}{c}{\multirow{6}[2]{*}{UKBB}} & Pix2pix \cite{Isola_2017_CVPR} & 32.77 & 0.925 & 19.30 & 0.674 & 18.89 & 0.653 \\
          & MMGAN \cite{8859286} & 32.72 & 0.927 & 19.16 & 0.669 & 18.96 & 0.669 \\
          & ResViT \cite{9758823} & 34.74 & 0.948 & 19.29 & 0.677 & 18.90 & 0.655 \\
          & UNETR \cite{Hatamizadeh_2022_WACV} & 36.55 & 0.964 & 19.31 & 0.680 & 18.94 & 0.653 \\
          & MTT-Net \cite{10268458} & 35.24 & 0.951 & 19.16 & 0.669 & 18.82 & 0.650 \\
          & SAGCNet (Ours) & \textbf{38.15} & \textbf{0.973} & \textbf{19.39} & \textbf{0.688} & \textbf{19.03} & \textbf{0.673} \\
    \midrule
    \multicolumn{1}{c}{\multirow{6}[2]{*}{MESA}} & Pix2pix \cite{Isola_2017_CVPR} & 28.83 & 0.811 & 22.25 & 0.742 & 21.72 & 0.738 \\
          & MMGAN \cite{8859286} & 28.62 & 0.810 & 22.25 & 0.740 & 21.71 & 0.734 \\
          & ResViT \cite{9758823} & 29.03 & 0.815 & 21.80 & 0.738 & 21.81 & 0.739 \\
          & UNETR \cite{Hatamizadeh_2022_WACV} & 33.23 & 0.884 & 22.21 & 0.764 & 21.93 & 0.747 \\
          & MTT-Net \cite{10268458} & 28.96 & 0.808 & 22.29 & 0.746 & 21.87 & 0.740 \\
          & SAGCNet (Ours) & \textbf{33.57} & \textbf{0.888} & \textbf{22.51} & \textbf{0.770} & \textbf{21.95} & \textbf{0.762} \\
    \midrule
    \multicolumn{1}{c}{\multirow{6}[2]{*}{ACDC}} & Pix2Pix \cite{Isola_2017_CVPR} & 25.57 & 0.758 & 18.98 & 0.625 & 18.47 & 0.613 \\
          & MMGAN \cite{8859286} & 25.68 & 0.764 & 18.96 & 0.632 & 18.50 & 0.619 \\
          & ResViT \cite{9758823} & 25.09 & 0.751 & 18.91 & 0.623 & 18.38 & 0.610 \\
          & UNETR \cite{Hatamizadeh_2022_WACV} & 29.30 & 0.862 & 19.03 & 0.631 & 18.60 & 0.624 \\
          & MTT-Net \cite{10268458} & 25.16 & 0.756 & 19.02 & 0.620 & 18.55 & 0.611 \\
          & SAGCNet (Ours) & \textbf{30.19} & \textbf{0.889} & \textbf{19.12} & \textbf{0.637} & \textbf{18.63} & \textbf{0.627} \\
    \bottomrule
    \end{tabular}%
  \label{Table1}%
\end{table}%

\par To further evaluate the performance, we also compare the quantitative results of SAGCNet with I2GAN \cite{XIA2021101812}, which is specifically designed for the single missing slice scenario (i.e., $P=1$). As illustrated in Table \ref{Table2}, our SAGCNet yields better performance across three metrics. It can be noticed that I2GAN is limited to scenarios where a single slice is missing. 

\begin{table}[h]
  \centering
  \footnotesize
  \setlength{\tabcolsep}{1.6mm}
  \renewcommand{\arraystretch}{0.9}
  \caption{Quantitative performance comparison of our SAGCNet and I2GAN \cite{XIA2021101812} on two datasets under single missing slice configuration. The best performance is in \textbf{bold}.}
    \begin{tabular}{lllll}
    \hline \textbf{Dataset} & \textbf{Method} & MAE $\downarrow$ & PSNR $\uparrow$ & SSIM $\uparrow$ \\
    \hline \multirow{2}{*}{ UKBB } & I2GAN \cite{XIA2021101812} & 0.0257 & 26.64 & 0.873 \\
    & Ours & $\mathbf{0.0110}$ & $\mathbf{38.19}$ & $\mathbf{0.978}$ \\
    \hline \multirow{2}{*}{ ACDC } & I2GAN \cite{XIA2021101812} & 0.0428 & 24.39 & 0.846 \\
    & Ours & $\mathbf{0.0304}$ & $\mathbf{30.86}$ & $\mathbf{0.895}$ \\
    \hline
    \end{tabular}
  \label{Table2}%
\end{table}

\begin{figure*}[t!]
\centerline{\includegraphics[width=29pc]{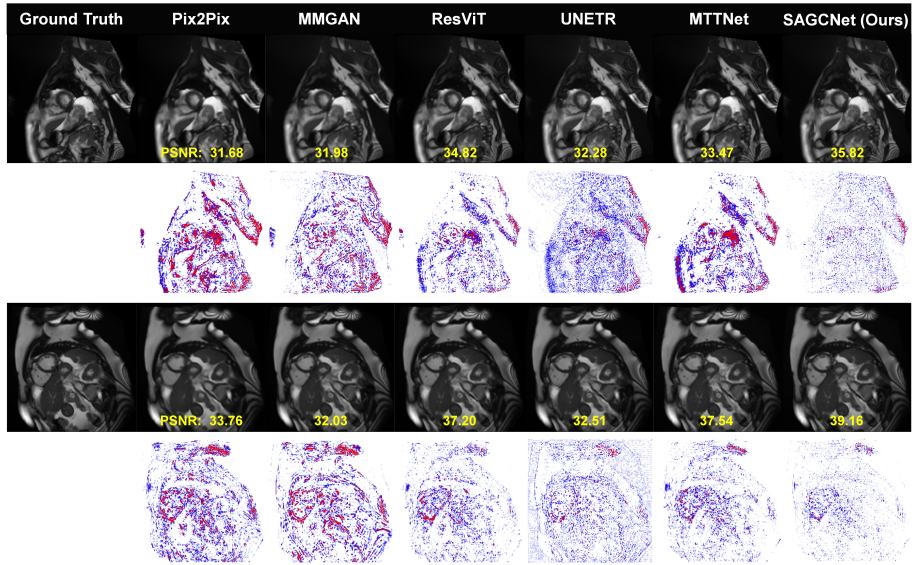}}
\caption{Qualitative results of all methods on the UKBB dataset. Every two rows, from top to bottom, denote the experimental results and the error maps, respectively. }
\label{Fig4}
\end{figure*}

\textbf{Qualitative Results.} The qualitative results of representative examples from the UKBB dataset are shown in Fig. \ref{Fig4}. It can be observed that baseline approaches tend to generate blurrier results that lack anatomical details and edges. In contrast, by modeling inter-slice dependencies and capturing global context information, our SAGCNet significantly reduces the red areas of the error maps of the synthesized images, particularly in anatomical structures and regions, indicating the capability of SAGCNet to produce results that are both visually realistic and closely aligned with the ground truth. 

\textbf{Ablation Studies.} To verify the effectiveness of each component in SAGCNet, ablation studies are conducted by constructing variant models. As shown in Table \ref{Table3}, each module contributes differently to the model performance on the UKBB and ACDC dataset, highlighting the positive impact of key components of SAGCNet in the missing slice imputation task. Furthermore, the VSA and VSGC modules exhibit the most significant performance improvements, further demonstrating the necessity and rationale of capturing inter-slice correlations and extracting valuable 3D information for volumetric data.

\begin{table}[!h]
  \setlength{\tabcolsep}{1.2mm}
  \renewcommand{\arraystretch}{0.6}
  \centering
  \footnotesize
  \caption{Quantitative performance comparison of ablation variants.}
  {
    \begin{tabular}{lccc||ccc}
    \toprule
    \multirow{3}[2]{*}{\textbf{Method}} & \multicolumn{6}{c}{\textbf{Dataset}} \\
    \multicolumn{1}{c}{} & \multicolumn{3}{c||}{UKBB} & \multicolumn{3}{c}{ACDC} \\
    \multicolumn{1}{c}{} & \multicolumn{1}{c}{MAE $\downarrow$} & \multicolumn{1}{c}{PSNR $\uparrow$} & \multicolumn{1}{c||}{SSIM $\uparrow$} & \multicolumn{1}{c}{MAE $\downarrow$} & \multicolumn{1}{c}{PSNR $\uparrow$} & \multicolumn{1}{c}{SSIM $\uparrow$} \\
    \midrule
    w/o VSA & 0.0130 & 37.05 & 0.967 & 0.0357 & 29.45 & 0.875 \\
    w/o VSGC & 0.0117 & 37.94 & 0.971 & 0.0329 & 30.08 & 0.879 \\
    w/o attribute-view & 0.0121 & 37.93 & 0.970 & 0.0331 & 29.98 & 0.875 \\
    w/o structure-view & 0.0119 & 37.91 & 0.968 & 0.0334 & 30.01 & 0.869 \\
    w/o $\mathcal L_{cl}$ & 0.0116 & 38.03 & 0.971 & 0.0328 & 30.12 & 0.881 \\
    SAGCNet & \textbf{0.0113} & \textbf{38.15} & \textbf{0.973} & \textbf{0.0326} & \textbf{30.19} & \textbf{0.889} \\
    \bottomrule
    \end{tabular}%
  \label{Table3}%
  }
\end{table}%

%% file: Sections/4_conclusion.tex
\section{Conclusion}
In this paper, we introduce SAGCNet for unified missing slice synthesis in volumetric CMR data. SAGCNet flexibly handles arbitrary sets of missing slices, reducing excessive reliance on complete volumetric data in real-world scenarios. By integrating inter-slice dependencies, we propose volumetric slice graphs that employ the graph structure to model CMR slices and impute absent data in a graph completion manner. To learn spatial-aware representations, we introduce the volumetric spatial adapter, empowering our model to adaptively exploit 3D spatial knowledge. Our extensive experiments demonstrate the superior effectiveness and robustness of SAGCNet compared with SOTA methods in both sufficient and limited slice data scenarios.

%% file: Paper-0554.bbl
\begin{thebibliography}{10}
\providecommand{\url}[1]{\texttt{#1}}
\providecommand{\urlprefix}{URL }
\providecommand{\doi}[1]{https://doi.org/#1}

\bibitem{10.5665/sleep.4732}
Chen, X., Wang, R., Zee, P., Lutsey, P.L., Javaheri, S., Alcántara, C., Jackson, C.L., Williams, M.A., Redline, S.: Racial/ethnic differences in sleep disturbances: The multi-ethnic study of atherosclerosis (mesa). Sleep  \textbf{38}(6),  877--888 (2015)

\bibitem{10.1007/978-3-031-72104-5_63}
Choo, K., Jun, Y., Yun, M., Hwang, S.J.: Slice-consistent 3d volumetric brain ct-to-mri translation with 2d brownian bridge diffusion model. In: MICCAI. pp. 657--667 (2024)

\bibitem{9758823}
Dalmaz, O., Yurt, M., Çukur, T.: Resvit: Residual vision transformers for multimodal medical image synthesis. IEEE Trans. Med. Imag.  \textbf{41}(10),  2598--2614 (2022)

\bibitem{Fang_2022_CVPR}
Fang, C., Wang, L., Zhang, D., Xu, J., Yuan, Y., Han, J.: Incremental cross-view mutual distillation for self-supervised medical ct synthesis. In: CVPR. pp. 20677--20686 (2022)

\bibitem{gasteiger2018predict}
Gasteiger, J., Bojchevski, A., G{\"u}nnemann, S.: Predict then propagate: Graph neural networks meet personalized pagerank. arXiv preprint arXiv:1810.05997  (2018)

\bibitem{Hatamizadeh_2022_WACV}
Hatamizadeh, A., Tang, Y., Nath, V., Yang, D., Myronenko, A., Landman, B., Roth, H.R., Xu, D.: Unetr: Transformers for 3d medical image segmentation. In: WACV. pp. 574--584 (2022)

\bibitem{Huo_Jin_Li_He_Yang_Wu_2023}
Huo, C., Jin, D., Li, Y., He, D., Yang, Y.B., Wu, L.: T2-gnn: Graph neural networks for graphs with incomplete features and structure via teacher-student distillation. AAAI  \textbf{37}(4),  4339--4346 (2023)

\bibitem{Isola_2017_CVPR}
Isola, P., Zhu, J.Y., Zhou, T., Efros, A.A.: Image-to-image translation with conditional adversarial networks. In: CVPR (2017)

\bibitem{10081095}
Liu, J., Pasumarthi, S., Duffy, B., Gong, E., Datta, K., Zaharchuk, G.: One model to synthesize them all: Multi-contrast multi-scale transformer for missing data imputation. IEEE Trans. Med. Imag.  \textbf{42}(9),  2577--2591 (2023)

\bibitem{10.1145/3580305.3599410}
Liu, Y., Ding, K., Wang, J., Lee, V., Liu, H., Pan, S.: Learning strong graph neural networks with weak information. In: KDD. p. 1559–1571. KDD '23 (2023)

\bibitem{NEURIPS2022_a92e9165}
Pan, J., Lin, Z., Zhu, X., Shao, J., Li, H.: St-adapter: Parameter-efficient image-to-video transfer learning. In: Proc. Adv. Neural Inform. Process. Syst. vol.~35, pp. 26462--26477 (2022)

\bibitem{Peng_2020_CVPR}
Peng, C., Lin, W.A., Liao, H., Chellappa, R., Zhou, S.K.: Saint: Spatially aware interpolation network for medical slice synthesis. In: CVPR (2020)

\bibitem{PENG2024102024}
Peng, X., Cheng, J., Tang, X., Zhang, B., Tu, W.: Multi-view graph imputation network. Inf. Fusion  \textbf{102},  102024 (2024)

\bibitem{PETERSEN20168}
Petersen, S.E., Matthews, P.M., Francis, J.M., Robson, M.D., Zemrak, F., Boubertakh, R., Young, A.A., Hudson, S., Weale, P., Garratt, S., Collins, R., Piechnik, S., Neubauer, S.: Uk biobank's cardiovascular magnetic resonance protocol. J. Cardiov. Magn. Reson.  \textbf{18}(1), ~8 (2016)

\bibitem{8859286}
Sharma, A., Hamarneh, G.: Missing mri pulse sequence synthesis using multi-modal generative adversarial network. IEEE Trans. Med. Imag.  \textbf{39}(4),  1170--1183 (2020)

\bibitem{velivckovic2017graph}
Veli{\v{c}}kovi{\'c}, P., Cucurull, G., Casanova, A., Romero, A., Lio, P., Bengio, Y.: Graph attention networks. arXiv preprint arXiv:1710.10903  (2017)

\bibitem{XIA2021101812}
Xia, Y., Zhang, L., Ravikumar, N., Attar, R., Piechnik, S.K., Neubauer, S., Petersen, S.E., Frangi, A.F.: Recovering from missing data in population imaging – cardiac mr image imputation via conditional generative adversarial nets. Med. Image Analys.  \textbf{67},  101812 (2021)

\bibitem{YURT2021101944}
Yurt, M., Dar, S.U., Erdem, A., Erdem, E., Oguz, K.K., Çukur, T.: mustgan: multi-stream generative adversarial networks for mr image synthesis. Med. Image Analys.  \textbf{70},  101944 (2021)

\bibitem{zhang2024automatic}
Zhang, L., Bronik, K., Piechnik, S.K., Lima, J.A., Neubauer, S., Petersen, S.E., Frangi, A.F.: Automatic plane pose estimation for cardiac left ventricle coverage estimation via deep adversarial regression network. IEEE Transactions on Artificial Intelligence  (2024)

\bibitem{zhang2019missing}
Zhang, L., Perea{\~n}ez, M., Bowles, C., Piechnik, S., Neubauer, S., Petersen, S., Frangi, A.: Missing slice imputation in population cmr imaging via conditional generative adversarial nets. In: MICCAI. pp. 651--659 (2019)

\bibitem{zhang2019unsupervised}
Zhang, L., Perea{\~n}ez, M., Bowles, C., Piechnik, S.K., Neubauer, S., Petersen, S.E., Frangi, A.F.: Unsupervised standard plane synthesis in population cine mri via cycle-consistent adversarial networks. In: MICCAI. pp. 660--668 (2019)

\bibitem{zhang2025diffuseg}
Zhang, L., Wu, F., Bronik, K., Papiez, B.W.: Diffuseg: Domain-driven diffusion for medical image segmentation. IEEE J. Biomed. Health Inform.  (2025)

\bibitem{ZHANG2024102381}
Zhang, T., Tan, T., Han, L., Wang, X., Gao, Y., {van Dijk}, J., Portaluri, A., Gonzalez-Huete, A., D’Angelo, A., Lu, C., Teuwen, J., Beets-Tan, R., Sun, Y., Mann, R.: Important-net: Integrated mri multi-parametric increment fusion generator with attention network for synthesizing absent data. Inf. Fusion  \textbf{108},  102381 (2024)

\bibitem{10589432}
Zhang, Y., Peng, C., Wang, Q., Song, D., Li, K., Kevin~Zhou, S.: Unified multi-modal image synthesis for missing modality imputation. IEEE Trans. Med. Imag.  \textbf{44}(1),  4--18 (2025)

\bibitem{10268458}
Zhong, L., Chen, Z., Shu, H., Zheng, K., Li, Y., Chen, W., Wu, Y., Ma, J., Feng, Q., Yang, W.: Multi-scale tokens-aware transformer network for multi-region and multi-sequence mr-to-ct synthesis in a single model. IEEE Trans. Med. Imag.  \textbf{43}(2),  794--806 (2024)

\bibitem{10183842}
Zhou, H.Y., Guo, J., Zhang, Y., Han, X., Yu, L., Wang, L., Yu, Y.: nnformer: Volumetric medical image segmentation via a 3d transformer. IEEE Trans. Image Process.  \textbf{32},  4036--4045 (2023)

\end{thebibliography}
